\documentclass[twocolumn]{aastex631} 

\usepackage{physics}
\usepackage{derivative}
\usepackage{bm}

\begin{document}

\title{Forming prominences accounting for partial ionisation effects}


\author[0000-0003-1862-7904]{V. Jer\v{c}i\'{c}}
\affiliation{NASA Goddard Space Flight Center, Greenbelt, MD, USA}

\author[0000-0001-5447-5456]{B. Popescu Braileanu}
\affiliation{Department of Physics and Technology, University of Bergen, Bergen, Norway}

\author[0000-0003-3544-2733]{R. Keppens}
\affiliation{Centre for mathematical Plasma-Astrophysics, Celestijnenlaan 200B, 3001 Leuven, KU Leuven, Belgium}

\begin{abstract}
One of the most striking structures in the solar atmosphere are prominences, predominantly coronal structures, with thermodynamic conditions that vary from chromospheric internally to the corona that surrounds them. These structures play an important role in the energy transfer between all layers of the atmosphere. Although mostly studied as a fully ionised plasma, prominences are, in fact, composed of partially ionised plasma. We do not yet fully understand the extent to which the two-fluid plasma-neutral properties play a role in the evolution of these coronal structures. In this work, we explore for the very first time how prominence formation and growth in a coronal loop evolves in a two-fluid setting. We used MPI-AMRVAC to study the evaporation-condensation process, where we consider radiative cooling, thermal conduction, and localised heating in a coronal loop in a fully stratified atmosphere. We report on the differences the two-fluid plasma brings into the prominence evolution, and more specifically in the period after the dynamic formation process finishes. Furthermore, we highlight the role it plays during the linear and the non-linear phases of the evolution. We find pronounced two-fluid effects in shocks that appear with the first complete condensation and confirm decoupling effects in the PCTR on the order of 100\,m\,s$^{-1}$, consistent with observations.
\end{abstract}


\section{Introduction} 
\label{sec:intro}

The photosphere and chromosphere are regions composed of partially ionised plasma, in which, as the density decreases with height, so does the coupling between charged and neutral elements \citep{Khomenko2012,Khomenko2017}. In addition to the photosphere and chromosphere, the corona also has regions where partially ionised plasma is important. High up in the solar atmosphere we find prominences. Prominences are plasma structures two orders of magnitude colder and denser than the corona. Their temperatures are on the order of 10$^4$\,K, and as such they consist of partially ionised plasma \citep{Ballester2018, Parenti2024}. Consequently, we expect the charges and neutrals in the conditions of prominences to behave, to an extent, independently. However, confirmation of the two-fluid properties is controversial. Measurement of the decoupling is not straightforward and is susceptible to misinterpretation. Multiple observational works have measured the decoupling in the velocities of ions and neutrals \citep{Khomenko2016,Wiehr2019,Wiehr2021,Zapior2022,Manrique2024}. \cite{Anan2017} report on different velocities of ions and neutrals, but unlike other authors, they do not conclude on the presence of two-fluid effects. They interpret the velocity differences as the result of observing different spectral lines of the same species; however, due to the different degree of optical thickness of each line, we observe different velocities \citep[as also demonstrated by][]{Zapior2022}.

In addition to observations, one can also model plasma-neutral mixtures numerically using a two-fluid description for the plasma and the neutrals. Two-fluid effects play an important role in the energy balance. Moreover, this interplay of charges and neutrals and their resulting dynamics plays a role in instabilities such as the Kelvin-Helmholtz and Rayleigh-Taylor instability, both of which were investigated through numerical simulations, e.g. \cite{Hillier2019} and \cite{Braileanu2021} respectively.

Most numerical work on the two-fluid prominence plasma has been done in relatively small simulation boxes (up to a few Mm in either direction). Small simulation domains allowed for high spatial resolution, which is crucial for determining and analysing partially ionised effects. There are also large-scale studies (with the domain extending tens of Mm in either direction) focused on prominences \citep{Terradas2015} or coronal rain \citep{Oliver2016,Martinez-Gomez2022}. However, many important effects, such as non-adiabatic effects that include the driving term for thermal instability that triggers prominence formation, were ignored in these simulations. In analogy with the early 1D hydro models of coronal loops forming prominences \citep{Antiochos1991,Karpen2001,Xia2011,Zhou2014,Zhang2012} we here initiate the counterpart of two-fluid 1D simulation in a large scale domain (of $\approx$112 Mm length of a coronal loop from footpoint to footpoint) of the formation of a cold (prominence) plasma. For the first time, in a two-fluid, fully stratified atmosphere, we include thermal conduction for charges and neutrals, radiative cooling, and localised heating. In Sec.~\ref{sec:equations_methods} we present the two-fluid equations and describe the numerical methods. In Sec.~\ref{sec:results} we present the results and describe the formation of the condensation and the influence of two-fluid properties on the evolution of the condensation. Finally, in Sect.~\ref{sec:disc_conc} we discuss the two-fluid effects and make concluding remarks.

\section{Two-fluid equations and numerical methods} 
\label{sec:equations_methods}

We report on 1D simulations performed with a state-of-the-art open-source MPI\ Adaptive Mesh Refinement Versatile Advection Code, ({\tt MPI-AMRVAC}\footnote{http://amrvac.org/}) \citep{Xia2018, Keppens2021, Braileanu2022, Keppens2023}. We simulate a coronal loop whose geometry has been used in other studies \citep{Antiochos1991,Karpen2001,Xia2011,Zhang2012,Zhang2013,Zhang2020} and is well described in \cite{Zhou2014}. In our case, the value of the parameters \citep[as mentioned in][]{Zhou2014} are the vertical leg with a length of $s_1=8$\,Mm, the radius of a quarter-circle shoulder $r=5$\,Mm, and the dip of depth $D=3$\,Mm. This results in a loop with a large horizontal extent and a very shallow dip. {\tt MPI-AMRVAC} offers the ability to have a changing cross section $A$, however, it is kept constant in our case. The height of the transition region is 2.719\,Mm and its width is 250\,km. We start with the temperature of the corona at 1\,MK, and the photosphere at 6000\,K. We set the total number density at the footpoints to be 5$\times10^{13}$\,cm$^{-3}$ with an ionisation degree of 0.05. These values were chosen in order to have appropriate coronal densities and pressures with as low a density of neutrals in the corona as possible (with the two-fluid module, a completely ionised corona is not numerically feasible, as exactly zero neutral density would translate to having vacuum conditions for neutrals).

We consider a hydrogen-only plasma composed of both neutral and charged species. To solve for the non-linear, compressible, two-fluid equations, we use a newly developed module in {\tt MPI-AMRVAC} \citep{Braileanu2022, Keppens2023}. The equations we solve are as follows,

\textbf{Continuity equations}
\begin{eqnarray}
\label{eq:continuity_n}
    && \pdv{\rho_n}{t} + \frac{1}{A(s)}\pdv*{\bigg(A(s)\rho_n v_n\bigg)}{s}= 0\,, \\
\label{eq:continuity_c}
    && \pdv{\rho_c}{t} + \frac{1}{A(s)}\pdv*{\bigg(A(s)\rho_c v_c\bigg)}{s}= 0\,.
\end{eqnarray}

\textbf{Momentum equations}
\begin{eqnarray}
\label{eq:momentum_n}
    && \rho_n \bigg[\pdv{v_n}{t} + v_n\pdv{v_n}{s}\bigg] =-\pdv{p_n}{s} -\rho_ng_{||} + R_n\,, \\
\label{eq:momentum_c}
    && \rho_c\bigg[\pdv{v_c}{t} + v_c\pdv{v_c}{s} \bigg] =-\pdv{p_c}{s} -\rho_cg_{||} - R_n\,.
\end{eqnarray}
\textbf{Energy equations}
\begin{eqnarray}
\label{eq:energy_n}
    &&\pdv{e_n^{tot}}{t} + \frac{1}{A(s)}\pdv*{}{s}\bigg[A(s)v_n(e^{tot}_n + p_n)\bigg] =  \rho_nv_ng_{||} + \, \nonumber \\
    && + \frac{1}{A(s)} \pdv*{}{s} \bigg(A(s){\kappa}_n \pdv{T_n}{s}\bigg) + M_n + H_n\,, \\
\label{eq:energy_c}
    &&\pdv{e_c^{tot}}{t} + \frac{1}{A(s)}\pdv*{}{s}\bigg[A(s)v_c (e^{tot}_c + p_c) \bigg] =  \rho_cv_cg_{||} + \, \nonumber \\
    && + \frac{1}{A(s)} \pdv*{}{s} \bigg(A(s){\kappa}_c \pdv{T_c}{s}\bigg) - n_c^2\Lambda(T_c) - M_n + H_c\,,
\end{eqnarray}
where we have,
\begin{eqnarray}
\label{eq:tot_e}
    &e^{tot}_c = e_c + \frac{1}{2}\rho_cv_c^2\,,\quad  \quad &e^{tot}_n = e_n + \frac{1}{2}\rho_nv_n^2\,, \\
\label{eq:internal_e}
    &e_c = p_c/(\gamma-1)\,, \quad &e_n = p_n/(\gamma-1)\,, \\
\label{eq:ideal_gas}
    &p_c =2 \rho_c T_c\,, \quad &p_n = \rho_n T_n\,.
\end{eqnarray}
Equations are written considering changes along $s$, which is the arc length along the fixed loop. For the case presented here, the cross-sectional area $A$ is constant with $s$. R$_n$ and $M_n$ are collisional terms, 
\begin{equation}
\label{eq:Rn}
    R_n = \alpha\rho_n\rho_c(v_c - v_n)\,.
\end{equation}
R$_n$ is a momentum collisional exchange term, where $\alpha$ is a collisional parameter defined as \citep[same as Eq.(A.3) in][in its dimensional form]{Braileanu2022},
\begin{equation}
\label{eq:alpha}
    \alpha = \frac{2}{m_H^{3/2}\sqrt{\pi}}\sqrt{k_BT_{cn}}\Sigma\,.
\end{equation}
Here, $T_{cn}$ is the average temperature of charges and neutrals. $\Sigma$ represents the collisional cross section between ions and neutrals and is taken to be 10$^{-15}$\,cm$^2$ \citep{Braileanu2022}. Furthermore, with $\alpha$ we can also express collisional frequencies between neutrals and ions, $\nu_{ni}=\alpha\rho_c$ and between ions and neutrals $\nu_{in}=\alpha\rho_n$. The collisional term in the energy Eq.~\ref{eq:energy_n} and~\ref{eq:energy_c}, $M_n$, is given by
\begin{eqnarray}
\label{eq:Mn}
    M_n &=& \frac{1}{2}\alpha\rho_n\rho_c(v^2_c - v^2_n) + \\
    && \frac{1}{\gamma-1}\alpha\rho_n\rho_c(T_c - T_n)   \, \nonumber ,
\end{eqnarray}
where $\gamma$ is equal to 5/3. Further details on the implementation of these equations are given in \cite{Braileanu2022} where they were presented for fully multi-dimensional settings where one would also need to handle the induction equation and the divergence constraint. Equations~\ref{eq:continuity_n}-\ref{eq:energy_c} are given in their conservative and dimensionless form (constants such as the Boltzmann constant, $k_B$, magnetic permeability, $\mu_0$, and hydrogen mass, $m_H$ are absorbed in units in these six equations). Unit normalisation values are $\Bar{n}=1\times10^9$\,cm$^{-3}$, $\Bar{T}=1\times10^6$\,K and $\Bar{L}=1\times10^9$\,cm. 

To allow the prominence to self-consistently form, we include thermal conduction (TC), radiative cooling (RC), and localised heating \citep{Antiochos1991}. Thermal conduction is isotropic for neutrals and anisotropic for charges (although this will not cause any difference in a 1D simulation). We use the Spitzer conductivity coefficient \citep{SpitzerHarm1953} for neutrals and charges, 
\begin{equation}
{\kappa}_l = \kappa_S T_l^{5/2} \,,\text{for }l=c,n
\label{eq:Spitzer_conductivity}
\end{equation}
with $\kappa_S=8\times10^{-7}$ erg\,cm$^{2}$\,s$^{-1}$\,K$^{-7/2}$. For the charges, it is clear that the field aligned collisional processes are dominated by electrons \citep{OrrallZirker1961}, and so for the charges this temperature dependent conductivity is fully justified. For neutrals, one may argue that this prescription does not match the expected hard-sphere collisional behaviour that is relevant in a neutral-only context. There have been multiple different studies on the calculations of transport coefficients in various conditions \citep{OrrallZirker1961, VranjesKrstic2013, Pavlov2017, Arber2023}. Despite of it all, the actual conduction coefficient for neutrals in a charge-neutral mixture is still a matter of debate. In order to approach this choice in a deliberate way, we performed two simulations that were meant to represent the more extreme possibilities of either having a temperature dependent conduction that is similar for ions and neutrals (as in the Eq.~\ref{eq:Spitzer_conductivity} above), or one where we kept the charge-specific conduction, Spitzer conduction, but where we completely turned off thermal conduction for neutrals. This allowed us to understand the potential differences we can expect if TC for neutrals is smaller than the Spitzer-Härm estimate. These two simulations behaved qualitatively similar, with details differing in the precise timing of the condensation formation, the shocks that are present in the early moments of the prominence formation, and in the profiles across the prominence-corona transition region. However, our findings on the degree of decoupling and the role of two-fluid effects were almost unmodified. As there is no consensus on the correct value of TC of neutrals, and as we have explored two opposite possibilities for it, we therefore opt to present the results from the simulation where both species have the same conduction prescription. We leave it to future research to further improve on the thermodynamical modelling and reach a consensus on diffusion processes in charge-neutral mixtures.

The RC curve is denoted by $\Lambda(T)$ and in this study we use the optically thin "SPEX" cooling curve \citep{Schure2009}. Further descriptions and comparisons of the effects of different cooling curves can be found in \cite{Hermans2021}. The $H_l$ in the energy equations is the localised heating \citep[see][]{Antiochos1999,Xia2011}. 
\begin{equation}
\label{eq:loc_heating}
        H_l = \begin{cases}
        0 & s < l_0 \,, \\
        E_l & l_0 \leq s < l_1 \,, \\
        E_l\exp(-(s-l_1)/h) & l_1 \leq s < L/2 \,, \\
        E_l\exp(-(L-l_1-s)/h) & L/2 \leq s < L-l_1 \,, \\
        E_l & L-l_1 \leq s < L-l_0 \,, \\
        0 & \text{elsewhere} \,,
        \end{cases}
    \end{equation}
where $l = c, n$, $L$ represents the length of the domain ($\approx$112\,Mm) and $l_0=1$\,Mm, $l_1=3$\,Mm, $h=5$\,Mm. The heating functions $H_c$ and $H_n$ both have the same form as presented in Eq.~\ref{eq:loc_heating}, the only difference is in the amplitude $E_l$. The amplitude of localised heating for charges is $E_c = 10^{-4}$ and for neutrals $E_n = 10^{-5}$\,erg\,cm$^{-3}$\,s$^{-1}$. The values we used, together with the fact that the heating of neutrals has a smaller value, was to limit the amount of neutrals that was pushed up into the corona as a result of this localised heating. We add a linear ramp function that grows during 18\,min when we activate localised heating. Immediately after the heating reaches its maximum value, we again decrease the heating to half its amplitude over a time of 9\,min. After that, the heating remains continuous during our simulations. Due to the relatively low value of neutrals in the corona and the strong gradients present at the TR, a two-fluid simulation is numerically more challenging than a single-fluid one in its ability to maintain a stratified background state for both species. To overcome this, we use the capability of {\tt AMRVAC} to split variables in their equilibrium, background values that do not evolve over time, and perturbations that evolve over time, $\rho=\rho_0+\rho_1$, $p=p_0+p_1$ \citep[see also][]{Yadav2022}. This capability, together with numerically relaxing to an energy equilibrium in {\tt AMRVAC}, puts our system in complete force and energy equilibrium with zero velocities in the system prior to the introduction of localised heating. This is also expected to hold in actual loop systems, prior to any condensation formation, and allows one to link linear theory on thermal instability mode development with actual loop evolutions \citep{Xia2011}.

The domain used here is 112\,Mm long and we use 6 levels of AMR, starting with 2240 cells on the base level. We use Harten-Lax-van-Leer (HLL) approximate Riemann solver \citep{Harten1983} for spatial discretisation in combination with a second order {\tt vanleer} flux limiter \citep{vanLeer1974}. A three-step time discretisation is used with an IMEX type time integrator\footnote[1]{See~\url{https://amrvac.org/md_doc_time_discretization.html} for more details}. The cooling term is handled exactly while the collisional coupling terms between charges and neutrals are treated implicitly. The thermal conduction is handled using super-time-stepping strategies \citep[for further discussion and references see][]{Keppens2023}. Due to the prescribed split of equilibrium and perturbed variables, the boundary conditions (referring to the perturbed values) are simply set to symmetric for densities, pressures and magnetic field, while the velocity is asymmetric.
    
This is the first time non-adiabatic effects (thermal conduction of both charges and neutrals, RC and localised heating) have been considered in a large-scale prominence simulation with partially ionised plasma and in a fully stratified atmosphere. We limit ourselves to a simulation without ionisation recombination (although it is already part of the two-fluid module of {\tt MPI-AMRVAC}), which is left for future studies. 

\section{Results} 
\label{sec:results}

\subsection{Simulations}
\label{sec:sims}
We performed two 1D simulations, one is the fully realistic two-fluid model of interest and the other is a single-fluid reference simulation. In the latter, we still have both charges and neutrals; however, we artificially increase the parameter $\alpha$ by three orders of magnitude compared to the self-consistent two-fluid simulation (where $\alpha$ is calculated according to Eq.~\ref{eq:alpha}). This is done to have a single-fluid-like simulation but still have matching initial conditions for the sake of comparison. We will refer to this other run as a single-fluid simulation throughout the paper. 

If the time scales of all physical processes in the domain are longer than the time scale of the ion-neutral collisions, then the single-fluid approximation is satisfactory. However, if they are near or shorter than the ion-neutral collision time scales, then a two-fluid solution is needed \citep{Zaqarashvili2011}. We can easily calculate the collisional frequencies by knowing the densities and the $\alpha$ parameter, as mentioned in Sec.~\ref{sec:equations_methods}. From that we can then calculate the mean free path between ions and neutrals and between neutrals and ions. This will depend on the characteristic velocity of the system and different examples can be found in the literature. For example, \cite{Braileanu2023} used the Alfv{\'e}n velocity while \cite{Braileanu2022} used a combination of the Alfv{\'e}n and sound speed. As the Alfv{\'e}n velocity is not really relevant for this 1D study, we use only the sound speed values,
\begin{eqnarray}
    \lambda_{in} = \frac{{c_s}_n}{\nu_{in}}\,, \\
    \lambda_{ni} = \frac{{c_s}_c}{\nu_{ni}}\,.
\end{eqnarray}
Our adaptive, high-resolution run is set to resolve these mean free paths over the whole coronal domain in the two-fluid run. At the moment of prominence formation $\lambda_{ni}$ in the corona is about 200\,km, while $\lambda_{in}$ is about 1000\,km. The size of our grid cells at the highest level of refinement is about 1.5\,km (which is the value reached in the TR and the PCTR). However, there is the small central, core part of the prominence, where $\lambda_{ni}$ becomes so small that we cannot resolve it. Where this resolution particularly matters is, for example, in the shocks that appear as a consequence of prominence formation \citep{Xia2011, Fang2015} (further mentioned in Sec.~\ref{subsec:formation}). The shocks at $t=113$\,min have a width of about 150\,km, which is significantly larger than our resolution, which at that time and place reaches its highest refinement. As the shocks have a width of similar values as $\lambda_{ni}$ we can argue that the two-fluid effects are resolved by the dynamic scales of our simulation and henceforth are important.

\subsection{Two-fluid prominence} 
\label{subsec:formation}

\begin{figure}
    \centering
    \includegraphics[width=\linewidth]{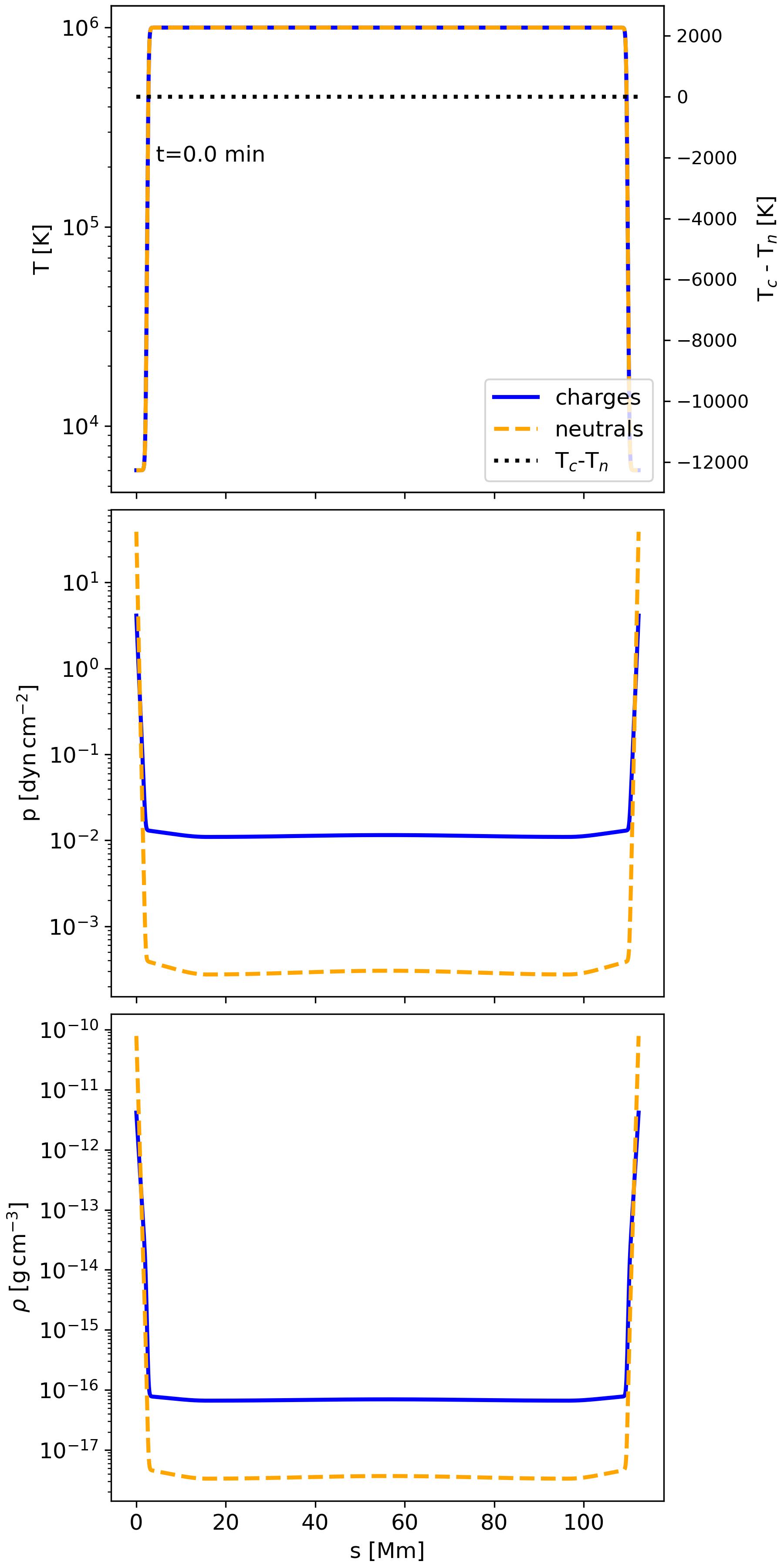}
    \caption{Initial state, at t=0 (before the localized heating is switched on).}
    \label{fig:initial_state}
\end{figure}
The initial equilibrium is shown in Fig.~\ref{fig:initial_state} where both the initial states of neutrals and charges are shown (orange dashed and blue full line, respectively) with velocities exactly equal to zero at $t=0$. We drive evaporation by symmetrically introducing localised heating at the footpoints of our loop system, which is maintained throughout the simulation (see Sec.~\ref{sec:equations_methods}). Heating increases the density inside the loop that locally perturbs the system enough that cooling begins at the middle of the domain due to the symmetry. RC overcomes localised heating and TC. As RC increases, it acts on the charges and the neutrals follow, considering the coupling is still strong enough. 
\begin{figure}
    \centering
    \includegraphics[width=\linewidth]{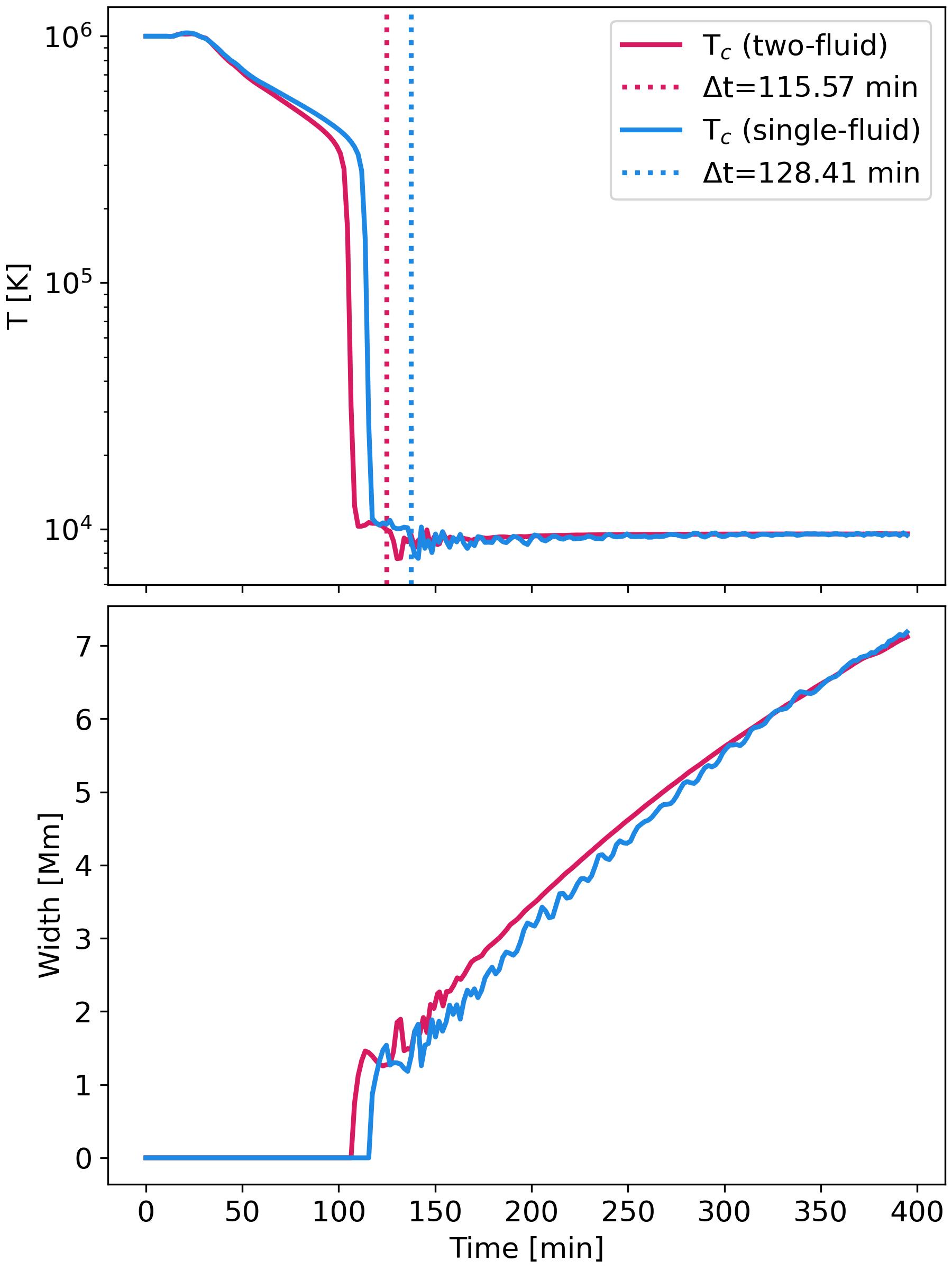}
    \caption{Top panel shows the temperature of charges at the middle of the domain (temperature of neutrals shows the exact same trend) and the bottom panel shows the width of the prominence (region in the corona where the density of charges is greater than 10$^9$\,cm$^{-3}$). The $\Delta t$ represents the time between the moment we turn on the localised heating and the moment the T$_c$ plotted here reaches 10$^4$\,K (represented by the dotted lines).}
    \label{fig:prominence_in_time}
\end{figure}
However, as the temperature starts to decrease, the coupling weakens - $\alpha$ drops and the collisional frequency decreases. Despite the initial conditions being the same and we include the same localised heating, the prominences in our two runs do not form at exactly the same time. From the upper panel of Fig.~\ref{fig:prominence_in_time} we see that the temperature in the two-fluid case drops to 10$^4$\,K about 13\,min earlier than in the single-fluid case. 
\begin{figure}
    \centering
    \includegraphics[width=\linewidth]{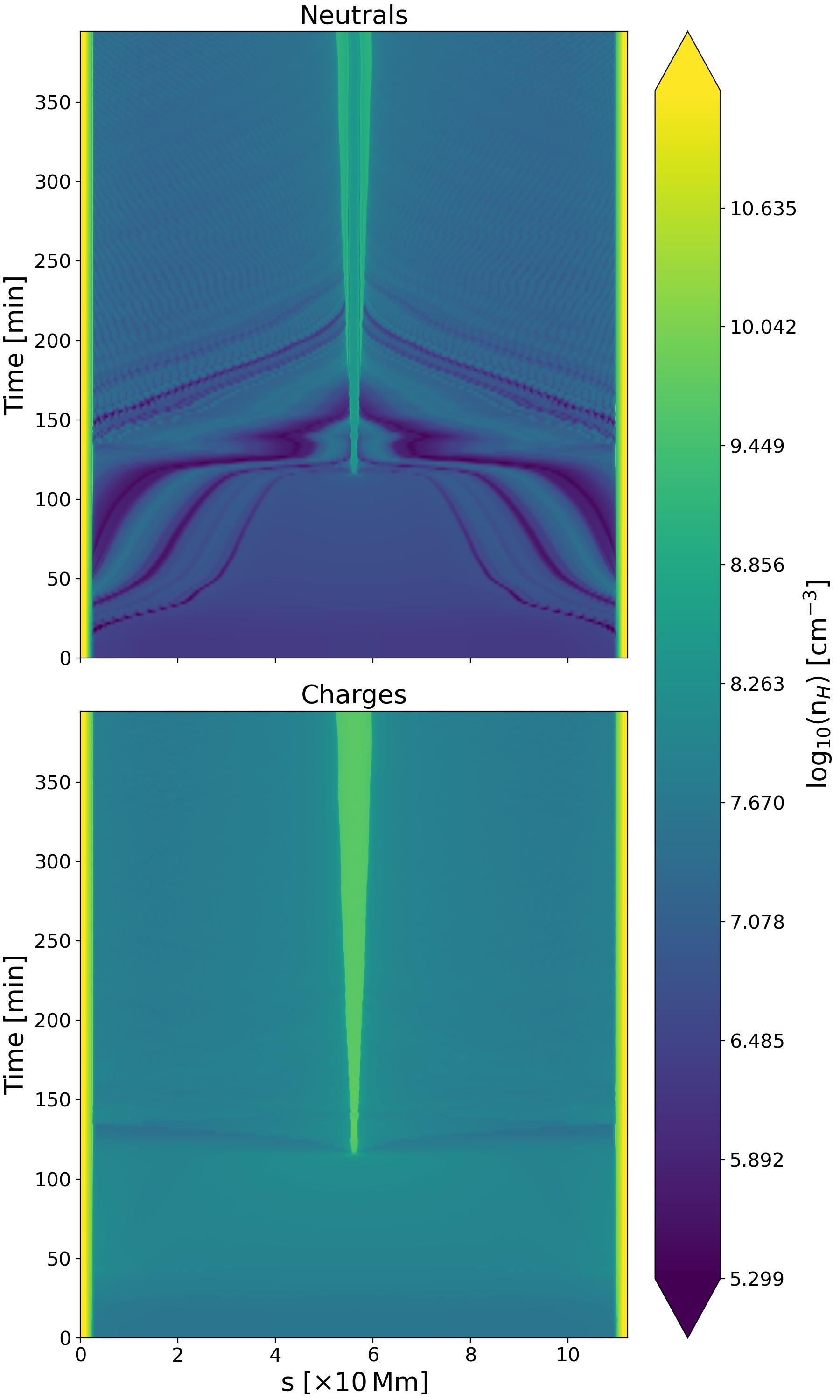}
    \caption{Small scale density perturbations due to propagating sound waves noticeable in the simulation where we artificially forced both fluids to be tightly coupled.}
    \label{fig:small_scale_structures}
\end{figure}
We notice small-scale structures in the form of sound waves that are abundant in the single-fluid simulation but completely absent in the two-fluid simulation. There have been multiple studies on the damping of waves caused by partially ionised plasma \citep{Soler2009,Soler2013,Zaqarashvili2011,Braileanu2019b} from which we already expect that the two-fluid simulation is less liable to sound wave perturbations. The small-scale structures, predominantly present in the neutrals of the single-fluid simulation, can be seen as density perturbations in Fig.~\ref{fig:small_scale_structures}. In the figure, they are clearly visible in the neutrals, and even though less visible, they are also present in the charges.

The ultimate local trigger of condensation is the onset of thermal instability. \cite{Braileanu2024} showed that the linear part of the growth of thermal instability depends on the collisional parameter $\alpha$. The higher $\alpha$, the slower the linear growth rate, meaning that the larger $\alpha$, in the case of a single fluid, influences the rate with which condensation forms.

\begin{figure*}
    \centering
    \includegraphics[width=\textwidth]{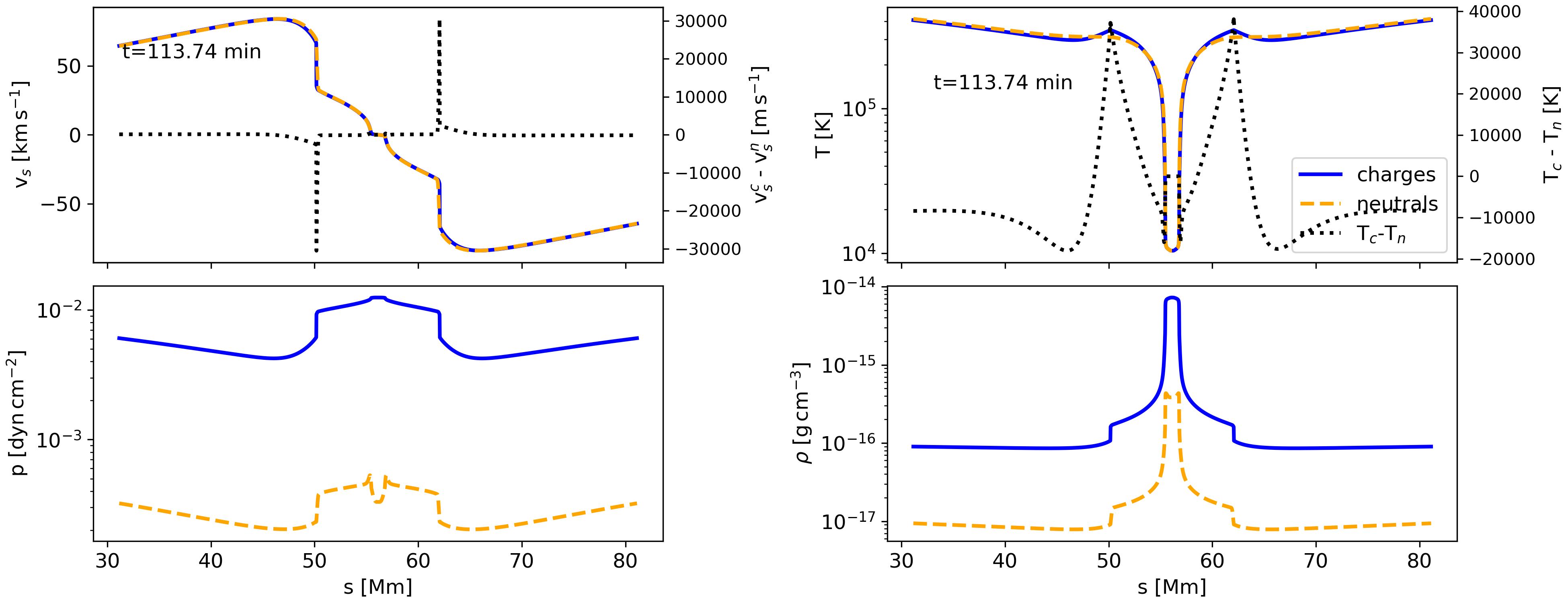}
    \caption{Parameters of the shocks that propagate after the condensation forms.}
    \label{fig:shock_param}
\end{figure*}

Returning to Fig.~\ref{fig:prominence_in_time} we see that the prominence reaches a temperature of 10$^4$\,K after about 115\,min for the two-fluid simulation and after 128\,min for the single-fluid one. The very early moments of formation are marked by a faster drop in temperature than the increase in density \citep[][and references therein]{Brughmans2022}. This results in the temperature outweighing the influence of density in the collisional frequencies. Consequently, as the RC grows stronger, the plasma grows colder and denser, and the decoupling between charges and neutrals is initiated. 
\begin{figure}
    \centering
    \includegraphics[width=\linewidth]{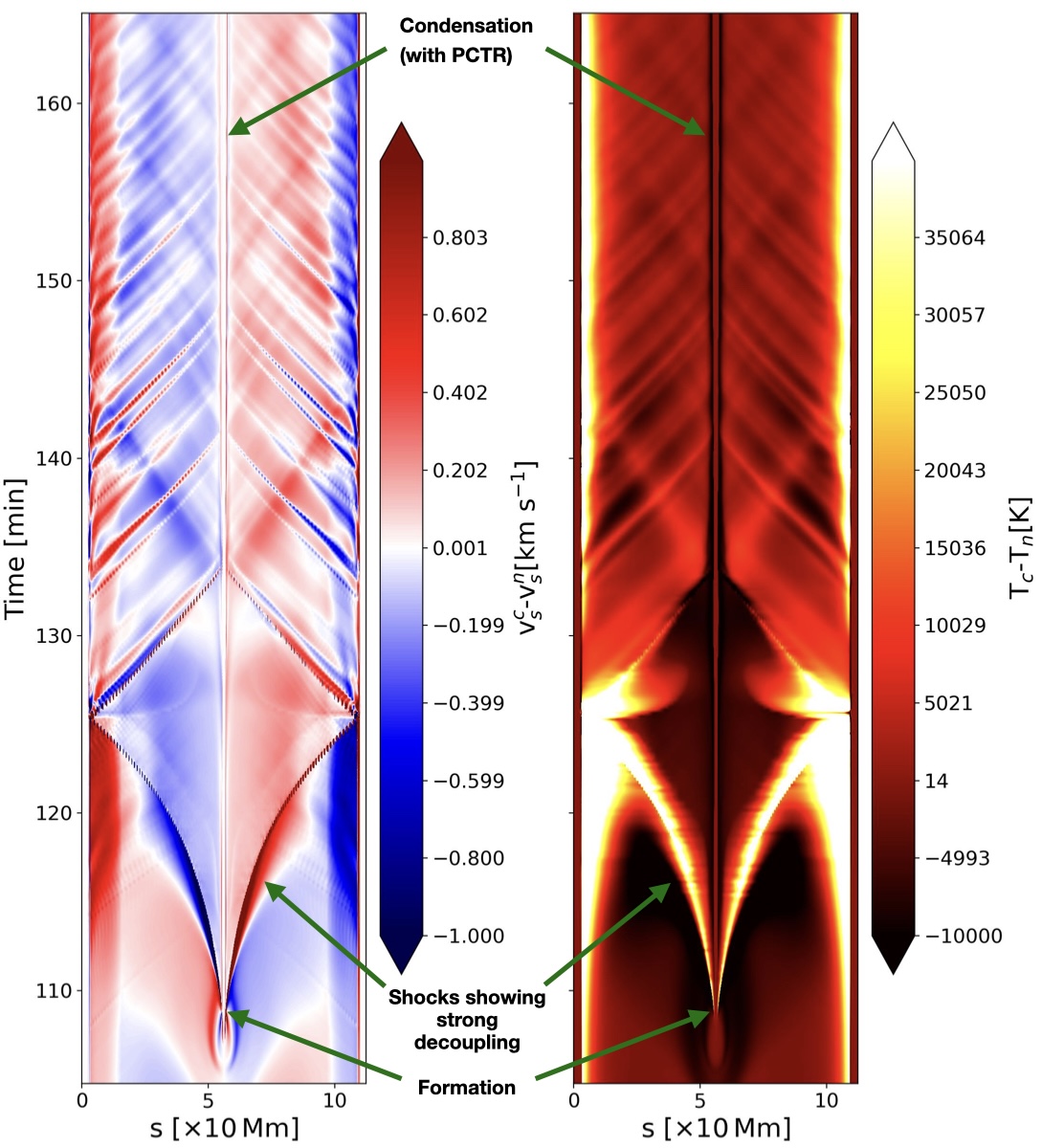}
    \caption{Velocity and temperature decoupling between charges and neutrals in the two-fluid simulation in the first hour after the prominence forms.}
    \label{fig:vT_differences}
\end{figure}
The velocity differences between charges and neutrals become ever more significant. In Fig.~\ref{fig:vT_differences} we show the velocity and temperature decoupling appearing in the two-fluid simulation for approximately the first hour after the condensation appears. The most significant decoupling (both in velocity and temperature) appears just after the prominence forms. Besides in the shocks that propagate away from the location of condensation, we also see large temperature and velocity decoupling in the evaporation flows coming from the footpoints and likewise being driven by the formation process. As the temperature drops below 20~000\,K we can already talk about a fully formed prominence. By that point, the drop in pressure that results from strong, localised cooling has already been compensated for by the incoming flows. The flows collide in the middle and reflect, creating shocks. The moment of formation together with the shocks resulting from it, are pointed out in the decoupling values shown in Fig.~\ref{fig:vT_differences}. Figure~\ref{fig:shock_param} shows velocity, temperature, pressure and density (from upper to lower and from left to right panel respectively) in the middle of the domain while the shocks are still clearly visible. These shocks represent another prominent feature in the prominence formation process. As other works already showed \citep{Hillier2016, Snow2019, Hillier_Snow2023}, shocks can be strongly influenced and modified by two-fluid plasma. From Fig.~\ref{fig:vT_differences} and~\ref{fig:shock_param} we see these shocks manifested in strong decoupling with temperatures differing even up to 40~000\,K and velocities up to a few tens of km\,s$^{-1}$. Furthermore, the shocks in the single-fluid simulation represent a sharp discontinuity. However, the shocks in the two-fluid simulation show a structure.
They are not merely a discontinuity, but rather a change in parameters that occurs over a certain spatial range \citep{Snow2019}. This is most clearly visible in Fig.~\ref{fig:shock_param} in the temperature profile. Even at a span of 50\,Mm we are showing here, both in the temperature but also in the temperature difference (dotted line) between charges and neutrals the shock structure is quite clear. As these shocks propagate, their width grows and eventually they dissipate. Besides the stark two-fluid effects, thermal conduction also has a strong influence on the dissipation of these shocks. They are, for the most part, well dissipated after about 180-190\,min into the simulation, and such large differences between the charge and neutral velocity and temperature are no longer seen. 

Further on, as our goal is to describe the stable phase in the prominence evolution, we focus on the part of the prominence evolution after the strong dynamics of the formation process has passed. After the shocks have dissipated we still clearly see that the prominence-corona transition region (PCTR) exhibits temperature and velocity decoupling. The values there are on the order of 10$^4$\,K and 100\,m\,s$^{-1}$, respectively. This results in extra source terms in the energy equation, such as frictional heating and thermal damping (which are also present in the mentioned shocks). This is the phase of the prominence evolution on which we will focus on in the following section.

\subsection{Heating}
\label{sec:heating}

\begin{figure*}
    \centering
    \includegraphics[width=\textwidth]{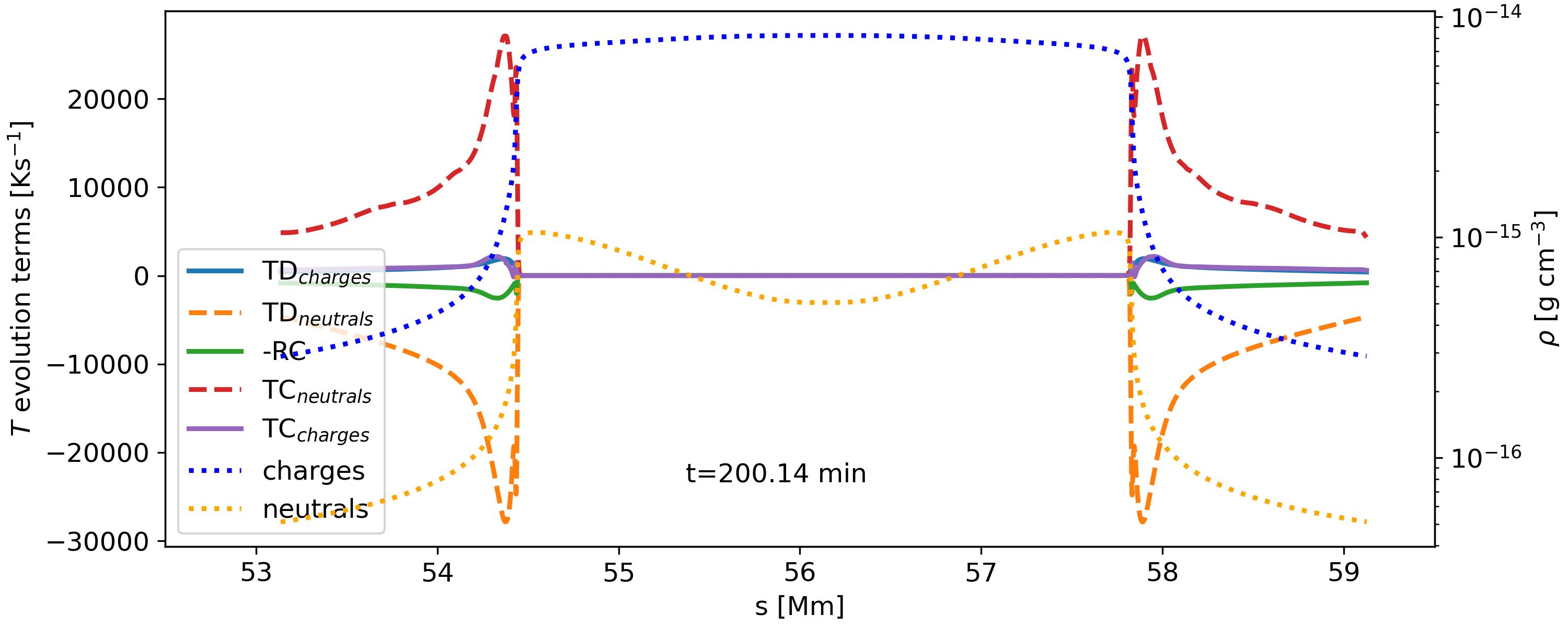}
    \caption{Temperature, $T$ evolution terms (full and dashed lines) at the prominence position (central part of the domain) at the moment when the shock waves resulting from the formation are mostly damped ($t=200$\,min). The dotted lines represent the density of charges and neutrals.}
    \label{fig:heating_structure}
\end{figure*}
To visualise the evolution of temperature, we rewrite the equations of energy (\ref{eq:energy_n} and~\ref{eq:energy_c}) using Eq.~\ref{eq:internal_e} together with the ideal gas closure, into the temperature equation. 
\begin{eqnarray}
    &&\pdv{T_n}{t} = -v_n\pdv{T_n}{s} + \frac{\gamma -1}{\rho_n}\big( -p_n \pdv{v_n}{s} +\, \nonumber  \\
    && TC_n + H_n \big) + FH_n + TD_n\,,
\end{eqnarray}
\begin{eqnarray}
    &&\pdv{T_c}{t} = -v_c\pdv{T_c}{s} + \frac{\gamma -1}{2\rho_c}\big( -p_c \pdv{v_c}{s} +\, \nonumber  \\
    && TC_c + RC + H_c \big) + FH_c + TD_c \,.
\end{eqnarray}
As before, $RC$ stands for radiative cooling and in the equation represents $-n_c^2 \Lambda(T_c)$. $H_c$ and $H_n$ are explained by Eq.~\ref{eq:loc_heating}. On the right-hand side, from the collisional term $M_n$ we get frictional heating (FH) and thermal damping (TD). In formula, we get
\begin{eqnarray}
\label{eq:FH_n}
    &&FH_n = \frac{1}{2}(v_c-v_n)^2\rho_c\alpha(\gamma-1)\,,\\
\label{eq:FH_c}
    &&FH_c = \frac{1}{4}(v_c-v_n)^2\rho_n\alpha(\gamma-1)\,,
\end{eqnarray}
of charges and neutrals, respectively. FH depends on the square of velocity difference, and as such it always plays the role of a source term. TD, on the other hand, depends on the thermal decoupling and plays the role of the source in one equation and a sink in the other
\begin{eqnarray}
\label{eq:TD_n}
    && TD_n = (T_c-T_n)\rho_c\alpha\,,\\
\label{eq:TD_c}
    && TD_c = \frac{1}{2}(T_n-T_c)\rho_n\alpha\,.
\end{eqnarray}
The two equations are always acting in a way to equalise the difference of temperatures between charges and neutrals. Figure~\ref{fig:heating_structure} shows the dominant terms in the temperature equation, mainly thermal damping, conduction, and cooling. Here we focus on the prominence way beyond the time shown in Fig.~\ref{fig:vT_differences}, which focuses on the shocks at the early stages of the formation. Figure~\ref{fig:heating_structure} shows the prominence after those shocks have passed. Optically thin cooling is the main driver of the decoupling, it initiates the velocity and temperature decoupling, which then results in FH and TD. Throughout the simulation, decoupling in the velocity and temperature is present at the PCTR. The two dominant terms, clearly seen in Fig.~\ref{fig:heating_structure}, are TD of neutrals (negative dashed orange line) and TC of neutrals (positive dashed red line). The two are close to an equilibrium, and they are the ones that dominate the temperature change of neutrals. In the case of charges, the dominant terms are the TC and TD (positive purple, and blue full lines, respectively) that oppose the influence of the RC (negative, green full line). As with neutrals, the terms are almost in equilibrium; however, the interplay results in the temperatures of both charges and neutrals continuously dropping at the PCTR. We also calculated and checked for the additional contributions of adiabatic compression \citep{Martinez-Gomez2022}, temperature advection, FH, and the influence of localised heating. All these terms are small compared to the ones shown in Fig.~\ref{fig:heating_structure} and do not show any significant contribution. The decrease in temperature with time at the PCTR is expected, as from Fig.~\ref{fig:prominence_in_time} we know that the condensation continuously grows in width. However, from the same figure we see that the single-fluid simulation catches up with the growth rate of the two-fluid condensation in the non-linear part of the evolution. We argue that although TD does not contribute to the heating of the whole system (see Eqs.~\ref{eq:TD_c} and~\ref{eq:TD_n}) it, nonetheless, influences individual temperatures of charges and neutrals. The cooling of charges is decreased as the RC now has to balance not only TC, but additionally the influence of TD.

As RC mostly vanishes at temperatures of 10$^4$\,K and lower, it does not act at the core of the prominence. Hence, there are no differences in the temperatures reached in the core of the prominence in the two-fluid and the single-fluid simulations. This may well be due to the adopted cooling curve prescription, and should be re-evaluated when transitions from optically thick to thin radiative processes are accounted for, as well as ionisation/recombination effects.

\section{Discussion and conclusion} 
\label{sec:disc_conc}

Here, for the first time we conducted a simulation of a partially ionised, large-scale prominence formation in a fully stratified atmosphere with non-adiabatic effects (optically thin RC, TC for charges and neutrals, and localised heating at the footpoints). 
We see that the two fluid effects are strongly present in the shocks that appear just after the prominence formation. After the shocks are dissipated, the two-fluid effects are still clearly observable at the PCTR. This is a result of the optically thin RC that affects the charges but not the neutrals, resulting in temperature and velocity decoupling. The decoupling at the PCTR is on the order of 100\,m\,s$^{-1}$ which is in good agreement with the observations of \cite{Khomenko2016}, and on the lower side of some of the other observations of velocity differences \citep{Stellmacher_Wiehr2017,Wiehr2019,Wiehr2021,Zapior2022,Manrique2024}. In all of these observational studies the ions are those with the larger velocity. In most cases, the cause argued for is the Lorentz force acting on the charges and not on the neutrals. As this simulation cannot account for the influence of the Lorentz force, it is expected that our values do not completely match the observed ones.

Despite the relatively simple domain and the fact that there is only field-aligned dynamics, the two-fluid simulation still shows distinct differences in comparison to the single-fluid one. Firstly, it shows a faster growth rate in the linear phase (dependence on the $\alpha$ parameter). Secondly, this changes later on, in the non-linear phase of the evolution, where the growth rate decreases in comparison to the single-fluid case. We show how relevant TC and TD really are. They are the dominant terms regulating changes at the PCTR. The study of \cite{Wojcik2020} demonstrated the contribution of collisional heating to balancing radiative and thermal energy losses in the corona. Similarly, we demonstrate the interplay of these energy terms and how it influences the heating of individual species, charges, and neutrals in the case of prominence evolution. 

Furthermore, even with all the additional physics we included, a more quantitative understanding will require even more physics; most importantly 2 or even 3 dimensionality and ionisation-recombination. Since we are not considering ionisation-recombination, we cannot trust our ionisation degree. This is important because the amount of neutrals can significantly influence the sources and sinks in the energy equation (which appear as a result of velocity and temperature decoupling). Furthermore, we showed how the particular optically thin RC drives the decoupling and the interplay of FH and TD with the rest of the energy terms. In order to better understand this, we still need to incorporate optically thick RC. It is still an open question as to how significant this optically thick RC is really in the energy balance. Nonetheless, this study offers invaluable insight into the interplay between charges and neutrals and represents a necessary step before we increase the complexity of our models. Furthermore, this opens up a whole new world of parameter study, and this analysis presents only the beginning. In the future, we plan to achieve a similar study in 2D where we expect the two-fluid effects to show a more significant contribution to our understanding of the dynamics across the magnetic field lines.

\begin{acknowledgments}
  The authors would like to thank the referee for his comments and feedback that helped to improve this work. VJ acknowledges funding from the Internal Funds KU Leuven and Research Foundation – Flanders FWO under project number 1161322N. VJ's research was further supported by an appointment to the NASA Postdoctoral Program at the Goddard Space Flight Center, administered by Oak Ridge Associated Universities under contract with NASA. BP has received funding from FWO, grant number 1232122N and European Union (ERC, DynaMIT, 101086985). RK is supported by the ERC Advanced Grant PROMINENT and an FWO grant G0B4521N and also acnowledges funding by C1 project UnderRadioSun C16/24/010 at KU Leuven, and FWO project G0B9923N Helioskill. This project has received funding from the European Research Council (ERC) under the European Union's Horizon 2020 research and innovation programme (grant agreement No. 833251 PROMINENT ERC-ADG 2018). Visualisations in this paper used \href{https://www.python.org/}{Python} and \href{https://yt-project.org/}{yt}. The resources and services used in this work were provided by VSC (Flemish Supercomputer Centre), funded by the Research Foundation - Flanders (FWO) and the Flemish Government.\\
  \textit{Facilities}: Vlaams Supercomputer Centrum\\
  \textit{Software}: MPI-AMRVAC \citep[][and references therein]{Keppens2023}, Python \citep{vanRossum}, Numpy \citep{harris2020array},
Matplotlib \citep{Hunter2007}, yt-project \citep{Turk2011} and \href{https://www.paraview.org/}{ParaView} \citep{ParaView}.
\end{acknowledgments}


\bibliography{references}{}

\begin{thebibliography}{}
\expandafter\ifx\csname natexlab\endcsname\relax\def\natexlab#1{#1}\fi
\providecommand{\url}[1]{\href{#1}{#1}}
\providecommand{\dodoi}[1]{doi:~\href{http://doi.org/#1}{\nolinkurl{#1}}}
\providecommand{\doeprint}[1]{\href{http://ascl.net/#1}{\nolinkurl{http://ascl.net/#1}}}
\providecommand{\doarXiv}[1]{\href{https://arxiv.org/abs/#1}{\nolinkurl{https://arxiv.org/abs/#1}}}

\bibitem[{Ahrens {et~al.}(2005)Ahrens, Geveci, \& Law}]{ParaView}
Ahrens, J., Geveci, B., \& Law, C. 2005, Visualization Handbook, ed. C.~D. Hansen \& C.~R. Johnson (Burlington, MA, USA: Elsevier Inc.), 717--731.
\newblock \url{https://www.sciencedirect.com/book/9780123875822/visualization-handbook}

\bibitem[{{Anan} {et~al.}(2017){Anan}, {Ichimoto}, \& {Hillier}}]{Anan2017}
{Anan}, T., {Ichimoto}, K., \& {Hillier}, A. 2017, \aap, 601, A103, \dodoi{10.1051/0004-6361/201629979}

\bibitem[{{Antiochos} \& {Klimchuk}(1991)}]{Antiochos1991}
{Antiochos}, S.~K., \& {Klimchuk}, J.~A. 1991, \apj, 378, 372, \dodoi{10.1086/170437}

\bibitem[{{Antiochos} {et~al.}(1999){Antiochos}, {MacNeice}, {Spicer}, \& {Klimchuk}}]{Antiochos1999}
{Antiochos}, S.~K., {MacNeice}, P.~J., {Spicer}, D.~S., \& {Klimchuk}, J.~A. 1999, \apj, 512, 985, \dodoi{10.1086/306804}

\bibitem[{{Arber} {et~al.}(2023){Arber}, {Goffrey}, \& {Ridgers}}]{Arber2023}
{Arber}, T.~D., {Goffrey}, T., \& {Ridgers}, C. 2023, Frontiers in Astronomy and Space Sciences, 10, 1155124, \dodoi{10.3389/fspas.2023.1155124}

\bibitem[{{Ballester} {et~al.}(2018){Ballester}, {Alexeev}, {Collados}, {Downes}, {Pfaff}, {Gilbert}, {Khodachenko}, {Khomenko}, {Shaikhislamov}, {Soler}, {V{\'a}zquez-Semadeni}, \& {Zaqarashvili}}]{Ballester2018}
{Ballester}, J.~L., {Alexeev}, I., {Collados}, M., {et~al.} 2018, \ssr, 214, 58, \dodoi{10.1007/s11214-018-0485-6}

\bibitem[{{Brughmans} {et~al.}(2022){Brughmans}, {Jenkins}, \& {Keppens}}]{Brughmans2022}
{Brughmans}, N., {Jenkins}, J.~M., \& {Keppens}, R. 2022, \aap, 668, A47, \dodoi{10.1051/0004-6361/202244071}

\bibitem[{{Fang} {et~al.}(2015){Fang}, {Xia}, {Keppens}, \& {Van Doorsselaere}}]{Fang2015}
{Fang}, X., {Xia}, C., {Keppens}, R., \& {Van Doorsselaere}, T. 2015, \apj, 807, 142, \dodoi{10.1088/0004-637X/807/2/142}

\bibitem[{{Gonz{\'a}lez Manrique} {et~al.}(2024){Gonz{\'a}lez Manrique}, {Khomenko}, {Collados}, {Kuckein}, {Felipe}, \& {G{\"o}m{\"o}ry}}]{Manrique2024}
{Gonz{\'a}lez Manrique}, S.~J., {Khomenko}, E., {Collados}, M., {et~al.} 2024, \aap, 681, A114, \dodoi{10.1051/0004-6361/202348119}

\bibitem[{Harris {et~al.}(2020)Harris, Millman, van~der Walt, Gommers, Virtanen, Cournapeau, Wieser, Taylor, Berg, Smith, Kern, Picus, Hoyer, van Kerkwijk, Brett, Haldane, del R{\'{i}}o, Wiebe, Peterson, G{\'{e}}rard-Marchant, Sheppard, Reddy, Weckesser, Abbasi, Gohlke, \& Oliphant}]{harris2020array}
Harris, C.~R., Millman, K.~J., van~der Walt, S.~J., {et~al.} 2020, Nature, 585, 357, \dodoi{10.1038/s41586-020-2649-2}

\bibitem[{Harten {et~al.}(1983)Harten, Lax, \& van Leer}]{Harten1983}
Harten, A., Lax, P., \& van Leer, B. 1983, SIAM Rev, 25, 35

\bibitem[{{Hermans} \& {Keppens}(2021)}]{Hermans2021}
{Hermans}, J., \& {Keppens}, R. 2021, \aap, 655, A36, \dodoi{10.1051/0004-6361/202140665}

\bibitem[{{Hillier}(2019)}]{Hillier2019}
{Hillier}, A. 2019, Physics of Plasmas, 26, 082902, \dodoi{10.1063/1.5103248}

\bibitem[{{Hillier} \& {Snow}(2023)}]{Hillier_Snow2023}
{Hillier}, A., \& {Snow}, B. 2023, Advances in Space Research, 71, 1962, \dodoi{10.1016/j.asr.2022.08.079}

\bibitem[{{Hillier} {et~al.}(2016){Hillier}, {Takasao}, \& {Nakamura}}]{Hillier2016}
{Hillier}, A., {Takasao}, S., \& {Nakamura}, N. 2016, \aap, 591, A112, \dodoi{10.1051/0004-6361/201628215}

\bibitem[{Hunter(2007)}]{Hunter2007}
Hunter, J.~D. 2007, Computing in Science \& Engineering, 9, 90, \dodoi{10.1109/MCSE.2007.55}

\bibitem[{{Karpen} {et~al.}(2001){Karpen}, {Antiochos}, {Hohensee}, {Klimchuk}, \& {MacNeice}}]{Karpen2001}
{Karpen}, J.~T., {Antiochos}, S.~K., {Hohensee}, M., {Klimchuk}, J.~A., \& {MacNeice}, P.~J. 2001, \apjl, 553, L85, \dodoi{10.1086/320497}

\bibitem[{{Keppens} {et~al.}(2023){Keppens}, {Popescu Braileanu}, {Zhou}, {Ruan}, {Xia}, {Guo}, {Claes}, \& {Bacchini}}]{Keppens2023}
{Keppens}, R., {Popescu Braileanu}, B., {Zhou}, Y., {et~al.} 2023, \aap, 673, A66, \dodoi{10.1051/0004-6361/202245359}

\bibitem[{Keppens {et~al.}(2021)Keppens, Teunissen, Xia, \& Porth}]{Keppens2021}
Keppens, R., Teunissen, J., Xia, C., \& Porth, O. 2021, Computers \& Mathematics with Applications, 81, 316, \dodoi{https://doi.org/10.1016/j.camwa.2020.03.023}

\bibitem[{{Khomenko}(2017)}]{Khomenko2017}
{Khomenko}, E. 2017, Plasma Physics and Controlled Fusion, 59, 014038, \dodoi{10.1088/0741-3335/59/1/014038}

\bibitem[{{Khomenko} \& {Collados}(2012)}]{Khomenko2012}
{Khomenko}, E., \& {Collados}, M. 2012, \apj, 747, 87, \dodoi{10.1088/0004-637X/747/2/87}

\bibitem[{{Khomenko} {et~al.}(2016){Khomenko}, {Collados}, \& {D{\'\i}az}}]{Khomenko2016}
{Khomenko}, E., {Collados}, M., \& {D{\'\i}az}, A.~J. 2016, \apj, 823, 132, \dodoi{10.3847/0004-637X/823/2/132}

\bibitem[{{Mart{\'\i}nez-G{\'o}mez} {et~al.}(2022){Mart{\'\i}nez-G{\'o}mez}, {Oliver}, {Khomenko}, \& {Collados}}]{Martinez-Gomez2022}
{Mart{\'\i}nez-G{\'o}mez}, D., {Oliver}, R., {Khomenko}, E., \& {Collados}, M. 2022, \apjl, 940, L47, \dodoi{10.3847/2041-8213/aca0a1}

\bibitem[{{Oliver} {et~al.}(2016){Oliver}, {Soler}, {Terradas}, \& {Zaqarashvili}}]{Oliver2016}
{Oliver}, R., {Soler}, R., {Terradas}, J., \& {Zaqarashvili}, T.~V. 2016, \apj, 818, 128, \dodoi{10.3847/0004-637X/818/2/128}

\bibitem[{{Orrall} \& {Zirker}(1961)}]{OrrallZirker1961}
{Orrall}, F.~Q., \& {Zirker}, J.~B. 1961, apj, 134, 63, \dodoi{10.1086/147128}

\bibitem[{{Parenti} {et~al.}(2024){Parenti}, {Luna}, \& {Ballester}}]{Parenti2024}
{Parenti}, S., {Luna}, M., \& {Ballester}, J.~L. 2024, Philosophical Transactions of the Royal Society of London Series A, 382, 20230225, \dodoi{10.1098/rsta.2023.0225}

\bibitem[{Pavlov(2017)}]{Pavlov2017}
Pavlov, A.~V. 2017, Journal of Geophysical Research: Space Physics, 122, 12,476, \dodoi{https://doi.org/10.1002/2017JA024397}

\bibitem[{{Popescu Braileanu} \& {Keppens}(2022)}]{Braileanu2022}
{Popescu Braileanu}, B., \& {Keppens}, R. 2022, \aap, 664, A55, \dodoi{10.1051/0004-6361/202243630}

\bibitem[{{Popescu Braileanu} \& {Keppens}(2023)}]{Braileanu2023}
---. 2023, \aap, 678, A66, \dodoi{10.1051/0004-6361/202346659}

\bibitem[{{Popescu Braileanu} \& {Keppens}(2024)}]{Braileanu2024}
---. 2024, Philosophical Transactions of the Royal Society of London Series A, 382, 20230217, \dodoi{10.1098/rsta.2023.0217}

\bibitem[{{Popescu Braileanu} {et~al.}(2019){Popescu Braileanu}, {Lukin}, {Khomenko}, \& {de Vicente}}]{Braileanu2019b}
{Popescu Braileanu}, B., {Lukin}, V.~S., {Khomenko}, E., \& {de Vicente}, {\'A}. 2019, \aap, 630, A79, \dodoi{10.1051/0004-6361/201935844}

\bibitem[{{Popescu Braileanu} {et~al.}(2021){Popescu Braileanu}, {Lukin}, {Khomenko}, \& {de Vicente}}]{Braileanu2021}
---. 2021, \aap, 650, A181, \dodoi{10.1051/0004-6361/202140425}

\bibitem[{{Schure} {et~al.}(2009){Schure}, {Kosenko}, {Kaastra}, {Keppens}, \& {Vink}}]{Schure2009}
{Schure}, K.~M., {Kosenko}, D., {Kaastra}, J.~S., {Keppens}, R., \& {Vink}, J. 2009, \aap, 508, 751, \dodoi{10.1051/0004-6361/200912495}

\bibitem[{{Snow} \& {Hillier}(2019)}]{Snow2019}
{Snow}, B., \& {Hillier}, A. 2019, \aap, 626, A46, \dodoi{10.1051/0004-6361/201935326}

\bibitem[{{Soler} {et~al.}(2013){Soler}, {D{\'\i}az}, {Ballester}, \& {Goossens}}]{Soler2013}
{Soler}, R., {D{\'\i}az}, A.~J., {Ballester}, J.~L., \& {Goossens}, M. 2013, \aap, 551, A86, \dodoi{10.1051/0004-6361/201220576}

\bibitem[{{Soler} {et~al.}(2009){Soler}, {Oliver}, \& {Ballester}}]{Soler2009}
{Soler}, R., {Oliver}, R., \& {Ballester}, J.~L. 2009, \apj, 699, 1553, \dodoi{10.1088/0004-637X/699/2/1553}

\bibitem[{{Spitzer} \& {H{\"a}rm}(1953)}]{SpitzerHarm1953}
{Spitzer}, L., \& {H{\"a}rm}, R. 1953, Physical Review, 89, 977, \dodoi{10.1103/PhysRev.89.977}

\bibitem[{{Stellmacher} \& {Wiehr}(2017)}]{Stellmacher_Wiehr2017}
{Stellmacher}, G., \& {Wiehr}, E. 2017, \solphys, 292, 83, \dodoi{10.1007/s11207-017-1103-6}

\bibitem[{{Terradas} {et~al.}(2015){Terradas}, {Soler}, {Oliver}, \& {Ballester}}]{Terradas2015}
{Terradas}, J., {Soler}, R., {Oliver}, R., \& {Ballester}, J.~L. 2015, \apjl, 802, L28, \dodoi{10.1088/2041-8205/802/2/L28}

\bibitem[{{Turk} {et~al.}(2011){Turk}, {Smith}, {Oishi}, {Skory}, {Skillman}, {Abel}, \& {Norman}}]{Turk2011}
{Turk}, M.~J., {Smith}, B.~D., {Oishi}, J.~S., {et~al.} 2011, \apjs, 192, 9, \dodoi{10.1088/0067-0049/192/1/9}

\bibitem[{{van Leer}(1974)}]{vanLeer1974}
{van Leer}, B. 1974, Journal of Computational Physics, 14, 361, \dodoi{https://doi.org/10.1016/0021-9991(74)90019-9}

\bibitem[{van Rossum(1995)}]{vanRossum}
van Rossum, G. 1995

\bibitem[{{Vranjes} \& {Krstic}(2013)}]{VranjesKrstic2013}
{Vranjes}, J., \& {Krstic}, P.~S. 2013, aap, 554, A22, \dodoi{10.1051/0004-6361/201220738}

\bibitem[{{Wiehr} {et~al.}(2021){Wiehr}, {Stellmacher}, {Balthasar}, \& {Bianda}}]{Wiehr2021}
{Wiehr}, E., {Stellmacher}, G., {Balthasar}, H., \& {Bianda}, M. 2021, \apj, 920, 47, \dodoi{10.3847/1538-4357/ac1791}

\bibitem[{{Wiehr} {et~al.}(2019){Wiehr}, {Stellmacher}, \& {Bianda}}]{Wiehr2019}
{Wiehr}, E., {Stellmacher}, G., \& {Bianda}, M. 2019, \apj, 873, 125, \dodoi{10.3847/1538-4357/ab04a4}

\bibitem[{{W{\'o}jcik} {et~al.}(2020){W{\'o}jcik}, {Ku{\'z}ma}, {Murawski}, \& {Musielak}}]{Wojcik2020}
{W{\'o}jcik}, D., {Ku{\'z}ma}, B., {Murawski}, K., \& {Musielak}, Z.~E. 2020, \aap, 635, A28, \dodoi{10.1051/0004-6361/201936938}

\bibitem[{{Xia} {et~al.}(2011){Xia}, {Chen}, {Keppens}, \& {van Marle}}]{Xia2011}
{Xia}, C., {Chen}, P.~F., {Keppens}, R., \& {van Marle}, A.~J. 2011, \apj, 737, 27, \dodoi{10.1088/0004-637X/737/1/27}

\bibitem[{{Xia} {et~al.}(2018){Xia}, {Teunissen}, {El Mellah}, {Chan{\'e}}, \& {Keppens}}]{Xia2018}
{Xia}, C., {Teunissen}, J., {El Mellah}, I., {Chan{\'e}}, E., \& {Keppens}, R. 2018, \apjs, 234, 30, \dodoi{10.3847/1538-4365/aaa6c8}

\bibitem[{{Yadav} {et~al.}(2022){Yadav}, {Keppens}, \& {Popescu Braileanu}}]{Yadav2022}
{Yadav}, N., {Keppens}, R., \& {Popescu Braileanu}, B. 2022, \aap, 660, A21, \dodoi{10.1051/0004-6361/202142688}

\bibitem[{{Zapi{\'o}r} {et~al.}(2022){Zapi{\'o}r}, {Heinzel}, \& {Khomenko}}]{Zapior2022}
{Zapi{\'o}r}, M., {Heinzel}, P., \& {Khomenko}, E. 2022, \apj, 934, 16, \dodoi{10.3847/1538-4357/ac778a}

\bibitem[{{Zaqarashvili} {et~al.}(2011){Zaqarashvili}, {Khodachenko}, \& {Rucker}}]{Zaqarashvili2011}
{Zaqarashvili}, T.~V., {Khodachenko}, M.~L., \& {Rucker}, H.~O. 2011, \aap, 529, A82, \dodoi{10.1051/0004-6361/201016326}

\bibitem[{{Zhang} {et~al.}(2012){Zhang}, {Chen}, {Xia}, \& {Keppens}}]{Zhang2012}
{Zhang}, Q.~M., {Chen}, P.~F., {Xia}, C., \& {Keppens}, R. 2012, \aap, 542, A52, \dodoi{10.1051/0004-6361/201218786}

\bibitem[{{Zhang} {et~al.}(2013){Zhang}, {Chen}, {Xia}, {Keppens}, \& {Ji}}]{Zhang2013}
{Zhang}, Q.~M., {Chen}, P.~F., {Xia}, C., {Keppens}, R., \& {Ji}, H.~S. 2013, \aap, 554, A124, \dodoi{10.1051/0004-6361/201220705}

\bibitem[{{Zhang} {et~al.}(2020){Zhang}, {Guo}, {Tam}, \& {Xu}}]{Zhang2020}
{Zhang}, Q.~M., {Guo}, J.~H., {Tam}, K.~V., \& {Xu}, A.~A. 2020, \aap, 635, A132, \dodoi{10.1051/0004-6361/201937291}

\bibitem[{{Zhou} {et~al.}(2014){Zhou}, {Chen}, {Zhang}, \& {Fang}}]{Zhou2014}
{Zhou}, Y.-H., {Chen}, P.-F., {Zhang}, Q.-M., \& {Fang}, C. 2014, Research in Astronomy and Astrophysics, 14, 581, \dodoi{10.1088/1674-4527/14/5/007}

\end{thebibliography}
\bibliographystyle{aasjournal}



\end{document}